\documentclass[3p,longtitle]{elsarticle}
\biboptions{sort&compress}
\usepackage[colorlinks=true, linkcolor=black, citecolor=blue, urlcolor=blue]{hyperref}
\usepackage{amsmath,amssymb}
\usepackage{siunitx}
\usepackage{booktabs}
\usepackage{textgreek}
\usepackage[caption=true]{subfig}
\usepackage[capitalise]{cleveref}

\newcommand\kr{\textsuperscript{83m}Kr}
\newcommand\rb{\textsuperscript{83}Rb}
\newcommand{\dd}[1]{\,\mathrm{d}#1}
\newcommand{\likelihood}{\mathcal{L}}
\newcommand{\krshell}[2]{#1\textsubscript{#2}}
\newcommand{\krline}[3]{\krshell{#1}{#2}-#3}

\ExplSyntaxOn
    \cs_undefine:N \c__siunitx_minus_tl
    \tl_const:Nn \c__siunitx_minus_tl {\scalebox{0.75}[1.0]{$-$}\thinspace}
\ExplSyntaxOff

\DeclareSIUnit\speedlight{\text{\ensuremath{c}}}
\DeclareSIUnit\cps{cps}
\DeclareSIUnit\ppm{ppm}

\begin{document}

\begin{frontmatter}

\title{High-resolution spectroscopy of gaseous \kr{} conversion electrons with the KATRIN experiment}


\address[a]{Technische Universit\"{a}t M\"{u}nchen, James-Franck-Str. 1, 85748 Garching, Germany}
\address[b]{IRFU, CEA, Universit\'{e} Paris-Saclay, 91191 Gif-sur-Yvette, France}
\address[c]{Helmholtz-Institut f\"{u}r Strahlen- und Kernphysik, Rheinische Friedrich-Wilhelms Universit\"{a}t Bonn, Nussallee 14-16, 53115 Bonn, Germany}
\address[d]{Karlsruhe Institute of Technology~(KIT), Institute of Experimental Particle Physics~(ETP), Wolfgang-Gaede-Str. 1, 76131 Karlsruhe, Germany}
\address[e]{Institut f\"{u}r Physik, Johannes-Gutenberg-Universit\"{a}t Mainz, 55099 Mainz, Germany}
\address[f]{Karlsruhe Institute of Technology~(KIT), Institute for Data Processing and Electronics~(IPE), Postfach 3640, 76021 Karlsruhe, Germany}
\address[g]{Institute for Nuclear Research of Russian Academy of Sciences, 60th October Anniversary Prospect 7a, 117312 Moscow, Russia}
\address[h]{Karlsruhe Institute of Technology~(KIT), Institute for Technical Physics~(ITeP), Postfach 3640, 76021 Karlsruhe, Germany}
\address[i]{Max-Planck-Institut f\"{u}r Kernphysik, Saupfercheckweg 1, 69117 Heidelberg, Germany}
\address[j]{Max-Planck-Institut f\"{u}r Physik, F\"{o}hringer Ring 6, 80805 M\"{u}nchen, Germany}
\address[k]{Karlsruhe Institute of Technology~(KIT), Institute for Nuclear Physics~(IKP), Postfach 3640, 76021 Karlsruhe, Germany}
\address[l]{Laboratory for Nuclear Science, Massachusetts Institute of Technology, 77 Massachusetts Ave, Cambridge, MA 02139, USA}
\address[m]{Center for Experimental Nuclear Physics and Astrophysics, and Dept.~of Physics, University of Washington, Seattle, WA 98195, USA}
\address[n]{Nuclear Physics Institute of the CAS, v.~v.~i., CZ-250 68 \v{R}e\v{z}, Czech Republic}
\address[o]{Institut f\"{u}r Kernphysik, Westf\"{a}lische Wilhelms-Universit\"{a}t M\"{u}nster, Wilhelm-Klemm-Str. 9, 48149 M\"{u}nster, Germany}
\address[p]{Department of Physics, Faculty of Mathematics und Natural Sciences, University of Wuppertal, Gauss-Str. 20, 42119 Wuppertal, Germany}
\address[q]{Department of Physics, Carnegie Mellon University, Pittsburgh, PA 15213, USA}
\address[r]{Universidad Complutense de Madrid, Instituto Pluridisciplinar, Paseo Juan XXIII, n\textsuperscript{\b{o}} 1, 28040 - Madrid, Spain}
\address[s]{Department of Physics and Astronomy, University of North Carolina, Chapel Hill, NC 27599, USA}
\address[t]{Triangle Universities Nuclear Laboratory, Durham, NC 27708, USA}
\address[u]{University of Applied Sciences~(HFD)~Fulda, Leipziger Str.~123, 36037 Fulda, Germany}
\address[v]{Department of Physics, Case Western Reserve University, Cleveland, OH 44106, USA}
\address[w]{Institute for Nuclear and Particle Astrophysics and Nuclear Science Division, Lawrence Berkeley National Laboratory, Berkeley, CA 94720, USA}
\address[x]{Institut f\"{u}r Physik, Humboldt-Universit\"{a}t zu Berlin, Newtonstr. 15, 12489 Berlin, Germany}

\fntext[fn1]{Also affiliated with Oak Ridge National Laboratory, Oak Ridge, TN 37831, USA}

\author[a,b]{K.~Altenm\"{u}ller}
\author[c]{M.~Arenz}
\author[d]{W.-J.~Baek}
\author[e]{M.~Beck}
\author[f]{A.~Beglarian}
\author[d]{J.~Behrens}
\author[f]{T.~Bergmann}
\author[g]{A.~Berlev}
\author[h]{U.~Besserer}
\author[i]{K.~Blaum}
% ...
\cortext[corr]{Corresponding author}
\ead{slezak@mpp.mpg.de}
\author[k]{M.~Steidl}
\author[o]{N.~Steinbrink}
\author[h]{M.~Sturm}
\author[n]{M.~Suchopar}
\author[h]{M.~Suesser}
\author[r]{H.~H.~Telle}
\author[q]{L.~A.~Thorne}
\author[k]{T.~Th\"{u}mmler}
\author[g]{N.~Titov}
\author[g]{I.~Tkachev}
\author[k]{N.~Trost}
\author[k]{K.~Valerius}
\author[n]{D.~V\'{e}nos}
\author[c]{R.~Vianden}
\author[q]{A.~P.~Vizcaya~Hern\'{a}ndez}
\author[f]{M.~Weber}
\author[o]{C.~Weinheimer}
\author[h]{S.~Welte}
\author[h]{J.~Wendel}
\author[s,t]{J.~F.~Wilkerson\fnref{1}}
\author[d]{J.~Wolf}
\author[f]{S.~W\"{u}stling}
\author[g]{S.~Zadoroghny}
\author[h]{G.~Zeller}

\begin{abstract}
In this work, we present the first spectroscopic measurements of conversion electrons originating from the decay of metastable gaseous \kr{} with the Karlsruhe Tritium Neutrino (KATRIN) experiment.
The results obtained in this calibration measurement represent a major commissioning milestone for the upcoming direct neutrino mass measurement with KATRIN.
The successful campaign demonstrates the functionalities of the full KATRIN beamline.
The KATRIN main spectrometer's excellent energy resolution of $\sim \SI{1}{\electronvolt}$ made it possible
to determine the narrow \krline{K}{}{32} and \krline{L}{3}{32} conversion electron line widths with an unprecedented precision of $\sim \SI{1}{\percent}$.
\end{abstract}

\begin{keyword}
neutrino mass \sep electrostatic spectrometer \sep calibration \sep conversion electrons
\end{keyword}

\end{frontmatter}

\section{Introduction}
The results of neutrino oscillation experiments have shown conclusively that neutrinos are massive particles \cite{Cleveland1998, Fukuda1998, Ahmad2002}.
As oscillations provide only information on the differences of the mass eigenvalues squared ($\Delta m^2$), the absolute neutrino mass scale has to be addressed by other means.
Complementary results related to the absolute neutrino mass scale are provided by cosmological observations \cite{Bennett2013, Ade2016}, neutrinoless double \textbeta{}-decay searches \cite{Agostini2017, Gando2016, Albert2014}, and direct measurements that utilize \textbeta{}-decays \cite{KATRIN2005, Formaggio2012b, Gastaldo2017}.
The direct measurements do not require any assumptions of the neutrino nature or mass model, and rely solely on kinematic considerations.
As pointed out by Fermi in 1934 \cite{Fermi1934}, a non-zero neutrino mass manifests itself as a distortion near the endpoint region of the \textbeta{}-electron energy spectrum.
While the experimental energy resolution is not good enough to resolve individual neutrino mass states, the observable extracted from the \textbeta{}-spectrum is the effective electron (anti)-neutrino mass squared.
It is an incoherent superposition of the mass eigenvalues, $m_\beta^2 = \sum_i |U_{ei}|^2 m_i^2$, where $U_{ei}$ are the Pontecorvo-Maki-Nakagawa-Sakata mixing matrix elements \cite{Otten2008}.

A particularly suitable isotope for direct neutrino mass measurement is tritium due to its low endpoint energy of about \SI{18.6}{\kilo\electronvolt} and favorable decay properties (super-allowed $1/2^+ \rightarrow 1/2^+$ transition).
As of today, only upper limits on $m_\beta$ have been obtained; the most stringent of which come from the experiments in Mainz with $m_\beta < \SI{2.3}{\electronvolt\per\square\speedlight}$ (\SI{95}{\percent} C.L.) \cite{Kraus2005} and Troitsk with $m_\beta < \SI{2.05}{\electronvolt\per\square\speedlight}$ (\SI{95}{\percent} C.L.) \cite{Aseev2011}, respectively.
The KATRIN (KArlsruhe TRItium Neutrino) experiment is a next-generation tritium \textbeta{}-decay experiment designed to search for $m_\beta$ with a sensitivity of \SI{0.2}{\electronvolt\per\square\speedlight} (\SI{90}{\percent} C.L.) and a $5\sigma$ discovery potential of $m_\beta = \SI{0.35}{\electronvolt\per\square\speedlight}$ \cite{KATRIN2005}.
It utilizes a highly luminous windowless gaseous tritium source and an electrostatic spectrometer with high resolution and large angular acceptance.

For calibration and investigations of systematic effects in absolute neutrino mass experiments involving tritium \cite{Robertson1991, Picard1992, Stoeffl1995, Formaggio2012b}, monoenergetic conversion electrons from the decay of the metastable isotope \kr{} are an important tool.
\kr{} provides conversion electron lines with energies up to \SI{32}{\kilo\electronvolt} \cite{Venos2018}.
The natural widths of these lines are comparable to the resolution of the KATRIN main spectrometer.
The isotope has a short half-life of \SI{1.83}{\hour} and can be introduced directly into the experimental apparatus without the risk of long-term radioactive contamination.

In this paper, we report on the results of gaseous \kr{} conversion electron measurements performed with the full KATRIN beamline during the pre-tritium commissioning phase.
We have obtained high-resolution spectra of conversion electron lines at the energies of \SIlist[]{17.8; 30.5; 32.1}{\kilo\electronvolt}.
This allowed us to assess the performance of the complete KATRIN setup over a broad range of energies and different natural line widths.
We also report on the relevant physical parameters of the conversion electron lines extracted from these spectra by means of a maximum likelihood analysis.
Earlier reports on the line widths \cite{PhDOstrick2008, Picard1992b} used a condensed source for the measurements which may be subject to a possible broadening of the lines due to surface effects.
In Ref.~\cite{PhDOstrick2008} the broadening was described in a general way by convolving the electron line shape with a Gaussian function whose width served as an additional unconstrained free parameter in the analysis.
In Ref.~\cite{Campbell2001}, generic uncertainty estimates are given for the recommended line widths: \SIrange{5}{10}{\percent} for the \krshell{K}{} shell and \SIrange{10}{30}{\percent} for the \krshell{L}{3} subshell.

\section{The KATRIN experiment}
\label{sec:katrin_experiment}
The KATRIN electron spectrometer operates as an integrating electrostatic filter with magnetic adiabatic collimation (MAC-E filter) \cite{Lobashev1985, Picard1992}.
During neutrino mass measurements, electrons are delivered via \textbeta{}-decays of molecular tritium in the windowless gaseous tritium source (WGTS) \cite{Bornschein2006}.
To prevent tritium from reaching the MAC-E filter where it would cause elevated background, differential (DPS) and cryogenic pumping sections (CPS),
together forming the electron transport section, are installed between the WGTS and the spectrometer \cite{Arenz2018A, Gil2010}.
Electrons with sufficient kinetic energy are transmitted through the pre- and main spectrometers and are eventually counted by a 148-segmented Si PIN-diode --- the focal plane detector (FPD) \cite{Amsbaugh2015}.
The MAC-E filter spectroscopy technique was successfully applied at previous direct neutrino mass experiments in Mainz \cite{Kraus2005} and Troitsk \cite{Aseev2011}.
Gaseous tritium sources were used in the Los Alamos National Laboratory experiment \cite{Robertson1991} and in Troitsk \cite{Aseev2011}.
For a detailed overview of the technical aspects of the KATRIN apparatus, the reader is referred to Ref.~\cite{Arenz2018A}.

In the KATRIN experiment, \kr{} sources are applied in three forms: gaseous, condensed, and implanted.
Their common attribute is the continuous generation of \kr{} from electron capture decay of its parent radionuclide \rb{}, which has a half-life of \SI{86.2}{\day}.
The parent half-life ensures a continuous supply of the short-lived \kr{} necessary for spectroscopy measurements with the MAC-E filter.
In all cases, physical, chemical, or mechanical means are deployed to ensure that the \rb{} itself does not leave its housing.
In this paper, we focus on the analysis of the measurements obtained with the gaseous \kr{} source (GKrS).
A description of the other sources can be found in Ref.~\cite{Arenz2018A}.

The basic principle of the GKrS is to introduce gaseous \kr{} into the WGTS.
Similar to tritium, it behaves as a spatially distributed isotropic source of electrons,
which allows for testing of the entire KATRIN setup in the same configuration as that during the neutrino mass measurement.
Even more importantly, as \kr{} and tritium can share the common volume in the source beam tube, the GKrS will allow us to study space-charge effects in the tritium plasma contributing to the source potential.
Unaccounted for potential variations within the source would effectively smear out the tritium \textbeta{}-spectrum,
leading to a systematic shift of the observed $m_\beta^2$ to more negative values \cite{PhDKuckert2016}.
The only difference to standard tritium operation is the higher WGTS beam tube temperature of $T = \SI{100}{\kelvin}$, instead of the default \SI{30}{\kelvin} (achieved using a dual-phase bath with argon instead of neon \cite{Arenz2018A}); this change prevents the freeze-out of \kr{} on the beam-tube walls.
The use of \kr{} in a gaseous source for space charge investigation was reported by the Troitsk group \cite{Belesev2008}.
The description of the dedicated \kr{} generator, used for the commissioning measurements described below, can be found in Ref.~\cite{Sentkerestiova2018}.
In contrast to the operation with tritium, \kr{} gas was left to expand freely in the beam tube and was pumped only by the cold inner surface of the CPS.

\section{Measurements}
The energy of the conversion electron --- emitted from a particular subshell of the \kr{} atom inside the WGTS --- with respect to the beam tube vacuum level is \cite{Venos2018}
\begin{equation}
\label{eq:electron_energy}
  E = E_\gamma + E_{\gamma,\,\mathrm{rec}} - E_{e,\,\mathrm{rec}} - E_{e,\,\mathrm{bin}},
\end{equation}
where $E_\gamma$ is the energy of the corresponding gamma ray, $E_{\gamma,\,\mathrm{rec}}$ is the recoil energy after gamma-ray emission, $E_{e,\,\mathrm{rec}}$ is the recoil energy after electron emission, and $E_{e,\,\mathrm{bin}}$ is the electron atomic binding energy.
To analyze the electron energy, the spectrometer is biased with respect to the grounded source tube by a negative retarding voltage $U$ thus creating an electrostatic barrier.
The electron passes the barrier when its energy $E$ is equal to or larger than the spectrometer vacuum level.
Denoting the source and spectrometer work functions as $\Phi_\mathrm{src}$ and $\Phi_\mathrm{spec}$, respectively, the transmission condition is
\begin{equation}
\label{eq:retarding_energy}
  E \geq qU - (\Phi_\mathrm{src} - \Phi_\mathrm{spec}),
\end{equation}
where $q < 0$ is the electron charge and $qU$ is the retarding energy.
Thus, the MAC-E filter measures an effective electron energy $qU$ that appears to be shifted from the expected kinetic energy $E$ by the work-function difference $\Phi_\mathrm{src} - \Phi_\mathrm{spec}$.

The MAC-E filter has a finite energy resolution $\Delta E$, which is defined for adiabatic transport of electrons by
\begin{equation}
\label{eq:energy_resolution}
  \Delta E = \frac{B_\mathrm{min}}{B_\mathrm{max}} \frac{\gamma+1}{2} E,
\end{equation}
where $B_\mathrm{min}$ is the minimal magnetic field at the center (the analyzing plane), $B_\mathrm{max}$ is the maximal magnetic field at the exit of the spectrometer, and $\gamma$ is the relativistic gamma-factor.
The magnetic field configuration of $B_\mathrm{min} = \SI{2.7e-4}{\tesla}$ and $B_\mathrm{max} = \SI{4.2}{\tesla}$ was set up such that an energy resolution of $\Delta E = \SI{1.17}{\electronvolt}$ at $E = \SI{17.8}{\kilo\electronvolt}$ was obtained.
With a source magnetic field of $B_S = \SI{2.52}{\tesla}$, the maximum electron acceptance angle was $\theta_\mathrm{max} = \arcsin\sqrt{B_S/B_\mathrm{max}} \approx \ang{51}$ and the accepted forward solid angle fraction was $\Delta \Omega / 2\pi \approx \SI{37}{\percent}$.

We have measured the zero-energy-loss peak of the \krshell{K}{}, \krshell{L}{3}, and the doublet \krshell{N}{2}, \krshell{N}{3} conversion electrons of the \SI{32}{\kilo\electronvolt} transition, denoted as \krline{K}{}{32}, \krline{L}{3}{32}, and \krline{N}{2,3}{32}.
The \krline{K}{}{32} line has an energy of \SI{17.82}{\kilo\electronvolt}, which is about \SI{750}{\electronvolt} below the tritium \textbeta{}-spectrum endpoint and can be used for calibrating the spectrometers in tritium \textbeta{}-decay measurements.
Its line width is about \SI{2.7}{\electronvolt} \cite{Venos2018}.
The \krline{L}{3}{32} line with an energy of \SI{30.47}{\kilo\electronvolt} has a line width of about \SI{1.2}{\electronvolt} and a $\sim$\num{1.5} times higher intensity.
In the KATRIN experiment this line is foreseen to be used for space charge investigations in the WGTS.
The close doublet \krline{N}{2,3}{32} has a lower intensity but a natural width that is much smaller than the spectrometer resolution.
There are no other strong lines above its energy of \SI{32.14}{\kilo\electronvolt}.
This is an important feature for an integrating spectrometer since this line is superimposed on the intrinsic spectrometer background only.
This doublet is essential in studying the MAC-E filter transmission function.

The \kr{} conversion electron integral energy spectra were obtained by changing the MAC-E filter retarding energy equidistantly in the region around the centroid of each line.
For different lines, the interval of the region ranged from about \SIrange{15}{25}{\electronvolt} with a typical step size between \SIlist{0.2; 0.5}{\electronvolt}.
The acquisition time at each voltage point was uniform for a given line with typical values between \SIlist{60; 150}{\second}.
Thus, the scanning time of a single spectrum was negligible with respect to the decay constant of the parent \rb{}.
For a given high-voltage setting, an average count rate was determined for each FPD pixel by summing all detected events in the region \SIrange{-3}{2}{\kilo\electronvolt} around the expected electron energy \cite{Arenz2018A}.
Integral spectra are obtained by plotting the count rate against the retarding energy.
A $\sim$ \SI{50}{\hertz} high-voltage ripple was present during the measurements\footnote{
  The active regulation system to counteract the high-voltage ripple \cite{Arenz2018A} was not in operation, but is foreseen to be used in future measurements.
}.
It is a near-sinusoidal signal with amplitudes of \SI{187}{\milli\volt} at \SI{-18}{\kilo\volt} and \SI{208}{\milli\volt} at \SI{-30}{\kilo\volt} \cite{Arenz2018B}.
The integrated rate of the measured lines over \num{137} operating detector pixels that have observed the \kr{} electrons\footnote{
  Due to a small misalignment of the setup, some FPD pixels were shadowed and could not observe \kr{} electrons \cite{Arenz2018A}.
} amounted to about \SI{4.1}{\kilo\cps} (\krline{K}{}{32}), \SI{6.7}{\kilo\cps} (\krline{L}{3}{32}), and \SI{0.16}{\kilo\cps} (\krline{N}{2,3}{32}), respectively.
Detector dead time is negligible at these low count rates \cite{Arenz2018A}.

\section{Analysis}
\subsection{Electron line shape}
A Lorentzian function is used to describe the shape of the conversion electron differential energy distribution:
\begin{equation}
\label{eq:lorentzian}
  L(E; A, E_0, \Gamma) = \frac{A}{\pi} \frac{\Gamma/2}{(E-E_0)^2 + \Gamma^2/4}.
\end{equation}
The parameters are the normalization factor $A$, the effective line position (centroid) $E_0$, and the line width (full width at half maximum) $\Gamma$.
To account for thermal Doppler broadening at the temperature $T$, the Lorentzian is convolved with a Gaussian function (normalized to one), yielding a Voigt function $V$.
The Gaussian part is considered to have a centroid of zero and a fixed width of $\sigma = \sqrt{EkT(\gamma+1)m/M}$, where $k$ is the Boltzmann constant, $m$ the electron mass, and $M$ the \kr{} mass.
The width $\sigma$ is thus in the range of \SIrange{46}{62}{\milli\electronvolt}.

To describe the MAC-E filter response to electrons, we consider the relativistic transmission function
\begin{equation}
\label{eq:transmission_function}
  T(E,qU) =
  \begin{cases}
    0, & E-qU < 0, \\
    \frac{1-\sqrt{1-\frac{E-qU}{E} \frac{2}{\gamma+1} \frac{B_S}{B_\mathrm{min}} }}{1-\sqrt{1-\frac{B_S}{B_\mathrm{max}}}}, & 0 \leq E-qU \leq \Delta E, \\
    1, & E-qU > \Delta E.
  \end{cases}
\end{equation}
In the presence of the high-voltage (HV) ripple described above, each electron experiences a different retarding potential according to the actual phase of the ripple signal.
The variation of the potential within the electron transport time is negligible.
The observed events include all possible phase values, leading effectively to a broadening of the transmission function.
This broadening is taken into account by convolving the transmission function with a digitized oscilloscope waveform of the ripple taken during the measurements.
The consequence is the effective shift of the onset of the transmission to a lower retarding energy $qU_\mathrm{min} < qU$.
Furthermore, the energy resolution is affected by synchrotron energy loss of the electrons on their way towards the spectrometer.
Using the particle-tracking simulation package Kassiopeia \cite{Furse2017}, it was determined that the energy loss would lead to a degradation of about \SIrange{30}{40}{\milli\electronvolt} in energy resolution between \SIlist{18; 30}{\kilo\electronvolt}.
Altogether we obtain a more accurate description of the transmission function $T'(E, qU)$ after making the aforementioned corrections.

The conversion electron integral line shape is calculated as
\begin{equation}
\label{eq:integral_spectrum}
  I(qU; A, E_0, \Gamma) = \int_{qU_\mathrm{min}}^{+\infty} V(E; A, E_0, \Gamma) \, T'(E, qU) \dd{E}.
\end{equation}
Each detector pixel observes a different minimal magnetic field $B_{\mathrm{min},i}$ and retarding potential offset $\Delta qU_i$ in the analyzing plane due to residual field inhomogeneities there.
These quantities were obtained with Kassiopeia \cite{Furse2017}.

\subsection{\krline{K}{}{32} and \krline{L}{3}{32} lines}
\begin{figure}
\centering
\subfloat[]{\includegraphics[width=0.5\textwidth]{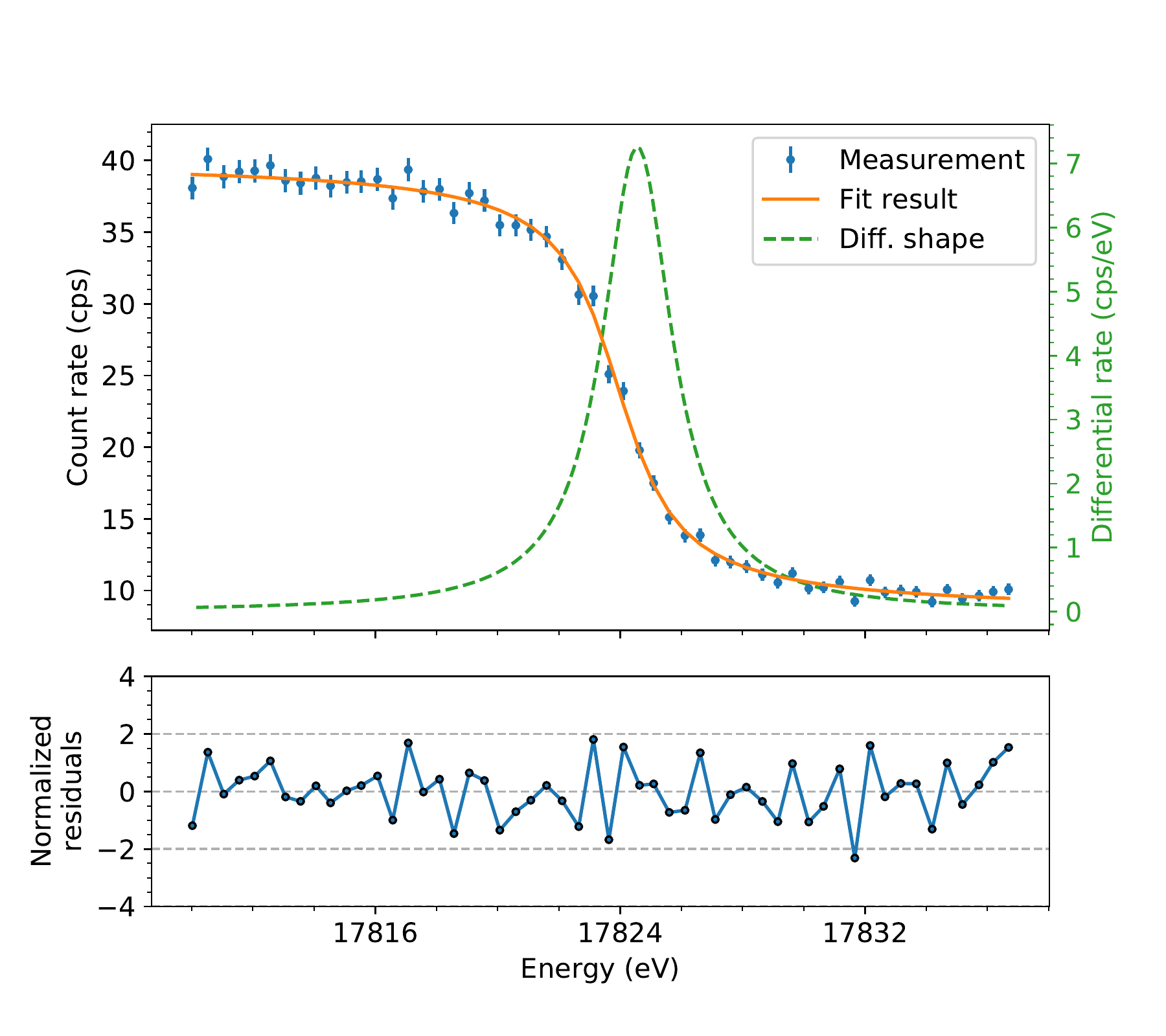}\label{fig:K-32}}
\subfloat[]{\includegraphics[width=0.5\textwidth]{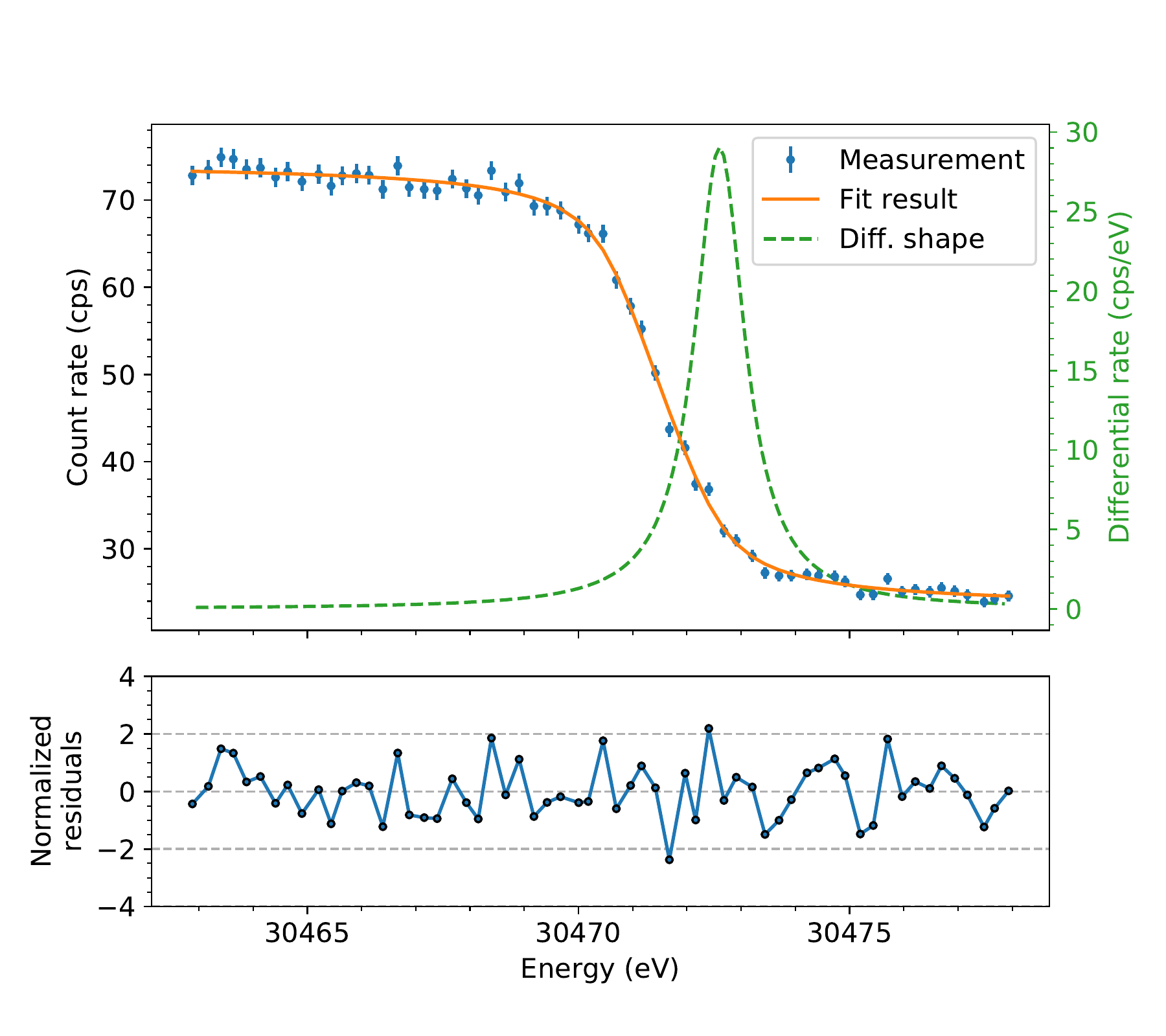}\label{fig:L3-32}}
\caption{The integral spectra of an active inner pixel and fit results of the \protect\subref{fig:K-32} \krline{K}{}{32} and \protect\subref{fig:L3-32} \krline{L}{3}{32} lines.
A negative shift corresponding to the potential offset in the analyzing plane $\Delta qU$ was added to the retarding energy $qU$.
The solid curve represents the integral spectrum model in \cref{eq:integral_spectrum} with the line shape parameters as obtained from the maximum likelihood analysis.
The dashed curve is the corresponding differential Lorentzian shape of the electron line, \cref{eq:lorentzian}.
The lower plots show the residuals of the fit normalized to the statistical uncertainty.
}
\label{fig:fits_K-32_L3-32}
\end{figure}

In the maximum likelihood analysis, we have assumed for the \krline{K}{}{32} and \krline{L}{3}{32} lines that the count rate $r=N/t$ follows the normal distribution,
and estimated its statistical uncertainty as $\sqrt{N}/t$, where $N$ is the measured number of counts per voltage step and $t$ is the acquisition time.
To account for the contributions of the higher-energy \kr{} lines and the intrinsic spectrometer background, a constant offset $C$ was added to the integral shape in \cref{eq:integral_spectrum} such that the fit model was $M = I(qU; A, E_0, \Gamma) + C$.
For each pixel, we performed $\chi^2(A, E_0, \Gamma, C)$ function minimization with the four variables as fit parameters.
Examples of the integral spectrum from an active inner pixel and the fit results for the \krline{K}{}{32} and \krline{L}{3}{32} lines are shown in \cref{fig:fits_K-32_L3-32}.
The fit model $M$ describes the observed spectrum without any residual structure.
The minimum chi-square per degree of freedom (dof) was $\chi^2_\mathrm{min}/\mathrm{dof} = \frac{47.28}{50} = 0.95$ and $\chi^2_\mathrm{min}/\mathrm{dof} = \frac{52.15}{57} = 0.92$ with the $p$-values of \num{0.58} and \num{0.66}, respectively.
The mean effective line position and line width from the results of all pixels, weighted by the reciprocal of squared statistical uncertainties obtained from the fit, are listed in \cref{tab:results}.

\subsection{\krline{N}{2,3}{32} doublet lines}
Since the \kr{} gas is very dilute, the vacancy left after electron emission from the outermost shell is expected to be long-lived.
Consequently, the natural line width is expected to be very narrow.
Owing to the spectrometer resolution of $\Delta E = \SI{2.13}{\electronvolt}$ at \SI{32}{\kilo\electronvolt} and the presence of the HV ripple, we approximated the narrow differential line shape by a \textdelta{}-function which is obtained in the limit $\Gamma \rightarrow 0$.
Thus, in this case, the fit model was based on a doublet of \textdelta{}-functions with $M = I(qU; A^{\mathrm{II}}, E_0^{\mathrm{II}}) + I(qU; A^{\mathrm{III}}, E_0^{\mathrm{III}}) + C$, where the upper indices II and III refer to \krline{N}{2}{32} and \krline{N}{3}{32}, respectively.

Due to the small number of counts at energies above the effective line position, we have assumed that the observed number of counts follows a Poisson distribution.
To break the degeneracy of the doublet parameters, a Gaussian penalty term was introduced into the likelihood function to restrict the difference of the effective line positions $\Delta E_0 = E_0^{\mathrm{III}}-E_0^{\mathrm{II}}$.
This difference is well known from optical spectroscopy measurements of electron binding energies to be $\Delta E_0^\mathrm{best} \, \pm \, \sigma(\Delta E_0) = \SI{0.670\pm0.014}{\electronvolt}$ \cite{Venos2018}.
The negative log-likelihood function is
\begin{equation}
\label{eq:likelihood_N}
  -\ln \likelihood (A^{\mathrm{II}}, E_0^{\mathrm{II}}, A^{\mathrm{III}}, E_0^{\mathrm{III}}, C)
  + \tfrac{1}{2} \left(\tfrac{\Delta E_0-\Delta E_0^\mathrm{best}}{\sigma(\Delta E_0)} \right)^2,
\end{equation}
where $\likelihood$ is the likelihood of the parameters in the argument given the observed counts.

An example of the spectrum from an active inner pixel and the fit result of the doublet are shown in \cref{fig:fit_N23-32}.
The fit residuals shown in the plot are defined as $res = \mathrm{sgn}(N-M_t)\sqrt{2\,[N \ln(N/M_t)-(N-M_t)]}$ with the model number of counts $M_t = Mt$.
We have estimated the $p$-value of the fit by means of a Monte Carlo study: toy measurements were generated from the best-fit model assuming Poisson distribution and corresponding negative log-likelihood function was minimized.
From the results of \num{e5} trials, the $p$-value was determined to be \num{0.44}.
The effective line positions of the \krline{N}{2,3}{32} doublet lines averaged over all pixels are listed in \cref{tab:results}.

\subsection{Systematic effects}
A significant systematic effect to be considered is the readout uncertainty of the HV system at the order of \SI{0.7}{\ppm} to \SI{0.9}{\ppm} in dependence on the line position \cite{Arenz2018B}.
Another contribution comes from the uncertainty of the transmission function width that we have estimated to be \SI{1}{\percent} \cite{PhDErhard2016}.
It is dominated by the uncertainties of the electric and magnetic field values in the analyzing plane and, to a lesser degree, by the variations in the path length of the electrons on their way through the source section, leading to different synchrotron energy losses.
We have also considered a conservative \SI{20}{\percent} uncertainty of the HV ripple \cite{Arenz2018B} experienced by the electrons at the analyzing plane.
The WGTS temperature and magnetic field stability were both one order of magnitude better than the design requirements \cite{Arenz2018A} and are negligible.
The combined systematic uncertainty of the effective line position and width from the considered contributions is listed in \cref{tab:results}.
The differences between the measured line positions can be used for a detailed cross-check of the high-voltage system calibration.
The results of this investigation are discussed in a separate paper \cite{Arenz2018B}.

\subsection{Expected and observed line position}
In order to relate the effective line position from \cref{tab:results} to the one expected from \cref{eq:electron_energy}, the difference of the source and spectrometer work functions, which was not determined beforehand, would have to be taken into account, see \cref{eq:retarding_energy}.
Besides, the expected line position is subject to a large systematic uncertainty of the 32-\si{\kilo\electronvolt} gamma-ray energy of \SI{0.5}{\electronvolt} \cite{Venos2018}.
The work-function difference and inaccuracy due to the gamma-ray energy uncertainty can contribute to a small constant shift between the expected and the effective line position.
As this shift is common for all lines of the \SI{32}{\kilo\electronvolt} transition, the expected and observed line positions can be compared while leaving the common offset free.

\begin{figure}
\centering
\subfloat[]{\includegraphics[width=0.5\textwidth]{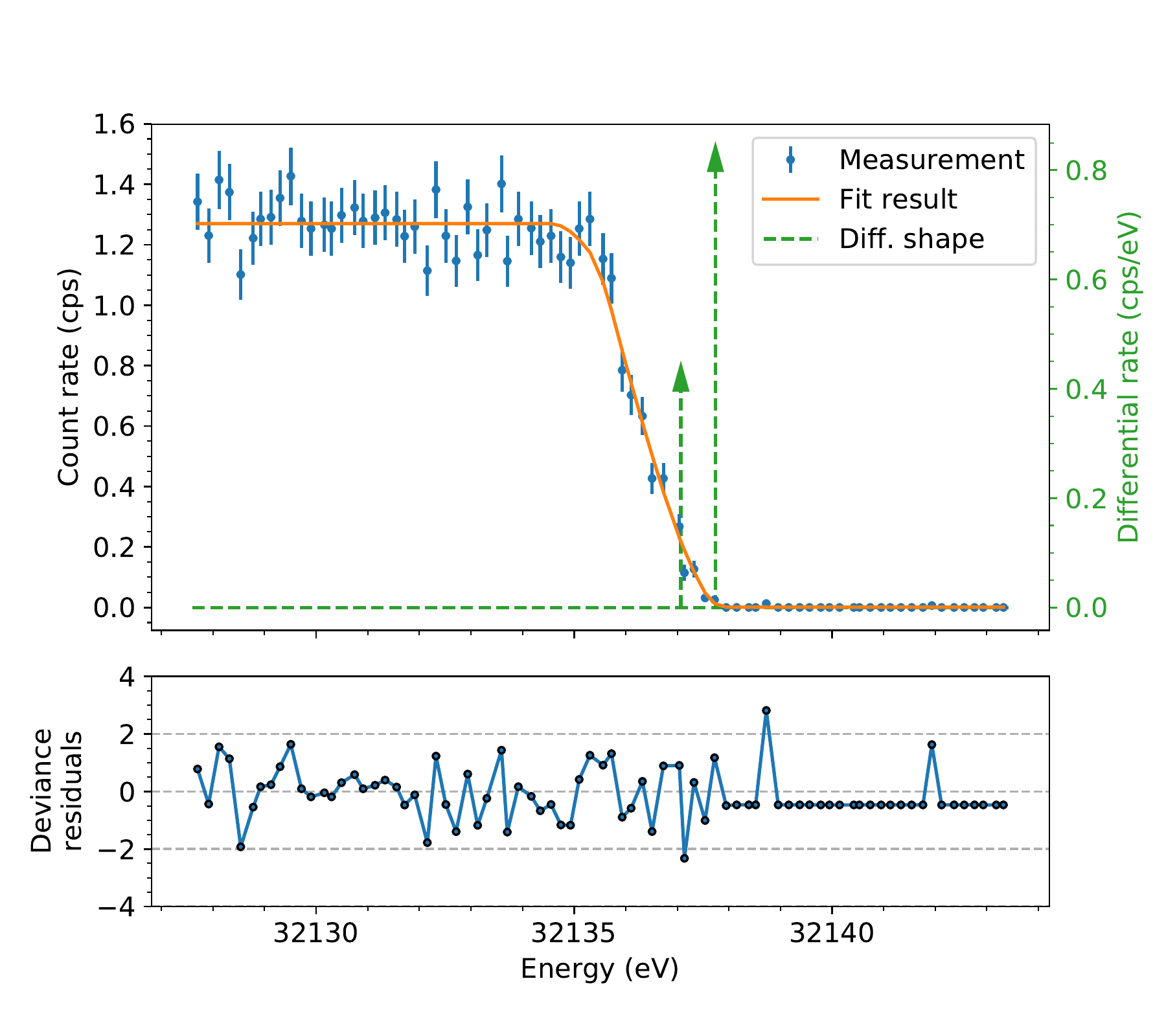}\label{fig:fit_N23-32}}
\subfloat[]{\includegraphics[width=0.5\textwidth]{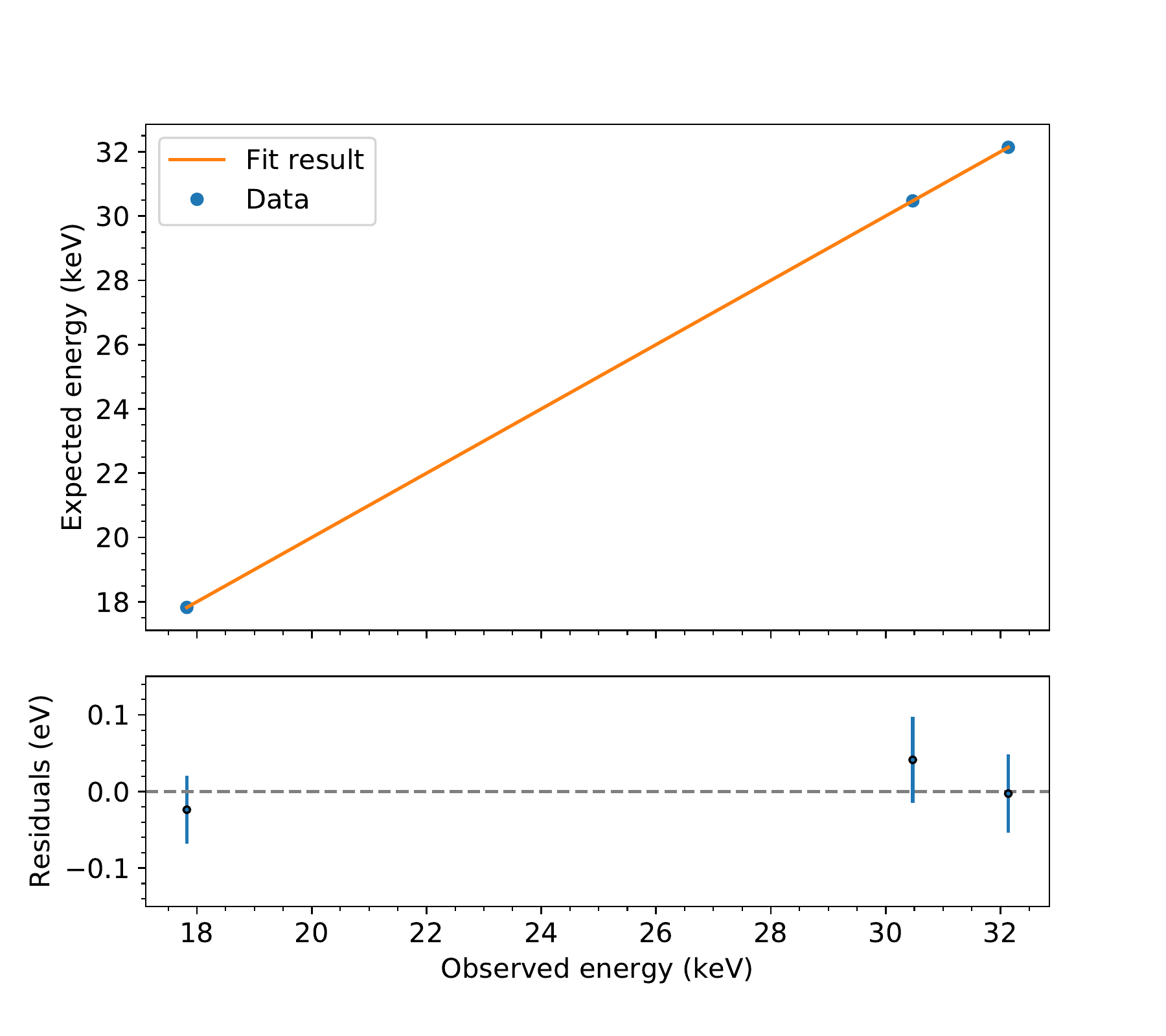}\label{fig:observed_vs_expected}}
\caption{\protect\subref{fig:fit_N23-32} The integral spectrum of an operating pixel and the fit results of the \krline{N}{2,3}{32} doublet.
The differential shape of both lines, for which zero natural width was assumed, is expressed by a \textdelta{}-function which is depicted here by an arrow.
The arrow height corresponds to the normalization factor obtained from the fit.
\protect\subref{fig:observed_vs_expected} A comparison of the effective and the expected line positions for the \krline{K}{}{32}, \krline{L}{3}{32}, and \krline{N}{3}{32} lines and a straight line fit to the data points assuming a fixed slope of one.
The uncertainties shown in the residual plot take into account contributions from both the expected and the effective line positions.
The common offset due to the gamma-ray energy uncertainty and the difference of work functions was left as a free parameter.}
\label{fig:fit_N-32_linearity}
\end{figure}

The expected line position is compared to the observed one in \cref{fig:observed_vs_expected}.
The statistical and systematic uncertainties of the observed line positions from \cref{tab:results} were added in quadrature.
The uncertainties of the expected line positions come from the uncertainties of the electron binding energies which are at the order of a few tenths of \si{\milli\electronvolt} \cite{Venos2018}.
Assuming a linear function with a fixed slope of one, the combined uncertainties showed in the residual plot of \cref{fig:observed_vs_expected} were obtained by adding the observed and expected line position uncertainties in quadrature.
The plot demonstrates the linearity of the KATRIN energy scale and the consistency of the \krline{N}{3}{32} effective line position from the doublet analysis of the \krline{N}{2,3}{32} region.
The uncertainty introduced due the common offset has no impact on the systematic uncertainty of the observed line width.

\section{Conclusion}
In summary, we have obtained high-resolution integral spectra of the \kr{} conversion electron lines with the KATRIN main spectrometer.
The spectra were analyzed by means of maximum likelihood analysis taking into account the relevant systematic effects, such as Doppler broadening, high-voltage ripple, synchrotron energy loss, and uncertainties of the electric and magnetic fields in the analyzing plane.
The results demonstrate the integrity of the full KATRIN beamline, an as-designed large angular acceptance and high energy resolution of the KATRIN main spectrometer,
good energy linearity over a range of \SI{14}{\kilo\electronvolt} and understanding of the observed spectra and hence the entire KATRIN apparatus.
Unprecedented precision of the \krline{K}{}{32} and \krline{L}{3}{32} line widths was achieved, thus improving over the existing results quantitatively
and serving as a reference for future calibration measurements at the KATRIN experiment and other direct neutrino mass experiments.
In a future measurement, an active regulation of the high-voltage system will be applied to reduce significantly the observed ripple which will further improve the spectrometer resolution.

\section*{Acknowledgements}
We acknowledge the support of Helmholtz Association (HGF), Ministry for Education and Research BMBF (5A17PDA, 05A17PM3, 05A17PX3, 05A17VK2, and 05A17WO3), Helmholtz Alliance for Astroparticle Physics (HAP), and Helmholtz Young Investigator Group (VH-NG-1055) in Germany; Ministry of Education, Youth and Sport (CANAM-LM2011019), cooperation with the JINR Dubna (3+3 grants) 2017–2019 in the Czech Republic; and the Department of Energy through grants DE-FG02-97ER41020, DE-FG02-94ER40818, DE-SC0004036, DE-FG02-97ER41033, DE-FG02-97ER41041, DE-AC02-05CH11231, DE-SC0011091, and DE-SC0019304 in the United States.

\bibliographystyle{elsarticle-num-names}
\bibliography{GKrSFirstResults}

\begin{table}
\centering
\caption{The best-fit parameters of the conversion electron lines averaged over individual pixels.
The \krline{N}{2,3}{32} spectrum was fitted using a doublet of \textdelta{}-functions the positions of which were constrained by the penalty term as described in the text.
In this case, no line width is given.
The average correlation coefficient of the \krline{N}{2,3}{32} centroids amounts to \num{0.98}.}
\label{tab:results}
\begin{tabular}{lccc}
\toprule
Line & Effective position $E_0$ (\si{\electronvolt}) & Width $\Gamma$ (\si{\electronvolt}) \\
\midrule
\krline{K}{}{32}  & \num[parse-numbers=false]{17824.576 \pm 0.005\textsubscript{stat} \pm 0.018\textsubscript{syst}} & \num[parse-numbers=false]{2.774 \pm 0.011\textsubscript{stat} \pm 0.005\textsubscript{syst}} \\[0.5em]
\krline{L}{3}{32} & \num[parse-numbers=false]{30472.604 \pm 0.003\textsubscript{stat} \pm 0.025\textsubscript{syst}} & \num[parse-numbers=false]{1.152 \pm 0.007\textsubscript{stat} \pm 0.013\textsubscript{syst}} \\[0.5em]
\krline{N}{2}{32} & \num[parse-numbers=false]{32137.098 \pm 0.016\textsubscript{stat} \pm 0.048\textsubscript{syst}} & -- \\[0.5em]
\krline{N}{3}{32} & \num[parse-numbers=false]{32137.758 \pm 0.015\textsubscript{stat} \pm 0.048\textsubscript{syst}} & -- \\
\bottomrule
\end{tabular}
\end{table}

\end{document}